\documentclass[preprint,12pt]{elsarticle}

\usepackage{graphicx}
\usepackage{subcaption}
\usepackage{amssymb}

\usepackage{lineno}
\usepackage{algorithm,algpseudocode} 
\usepackage{enumitem}	 
\usepackage{amsmath}	 
\usepackage{natbib} 	 
\usepackage[a4paper,  margin=1in]{geometry}	 

\journal{Transportation Research Part C: Emerging Tech.}

\begin{document}

\begin{frontmatter}


\title{Optimizing Signalized Intersections Performance under Conventional and Automated Vehicles Traffic}



\author[label1]{Mahmoud Pourmehrab\corref{cor1}}
\author[label1]{Lily Elefteriadou}
\author[label2]{Sanjay Ranka}
\author[label1]{Marilo Martin-Gasulla}

\cortext[cor1]{Corresponding author, e-mail: mpourmehrab@ufl.edu}
\address[label1]{University of Florida, 365 Weil Hall, PO Box 116580, Gainesville, FL 32611, United States}
\address[label2]{Department of CISE, University of Florida, Gainesville, FL 32611, United States}

\begin{abstract}
Automated vehicles, or AVs (i.e. those that have the ability to operate without a driver and can communicate with the infrastructure) may transform the transportation system. This study develops and simulates an algorithm that can optimize signal control simultaneously with the AV trajectories under undersaturated traffic flow of AV and conventional vehicles.  This proposed Intelligent Intersection Control System (IICS) operates based on real-time collected arrival data at detection ranges around the center of the intersection. Parallel to detecting arrivals, the optimized trajectories and signal control parameters are transmitted to AVs and the signal controller to be implemented. Simulation experiments using the proposed IICS algorithm successfully prevented queue formation up to undersaturated condition. Comparison of the algorithm to operations with conventional actuated control shows 38 \textendash \ 52\% reduction in average travel time compared to conventional signal control.
\end{abstract}


\end{frontmatter}


\section{Introduction}
\label{S:1}

\begin{figure}[h]
\centering
\includegraphics[width=0.6\linewidth]{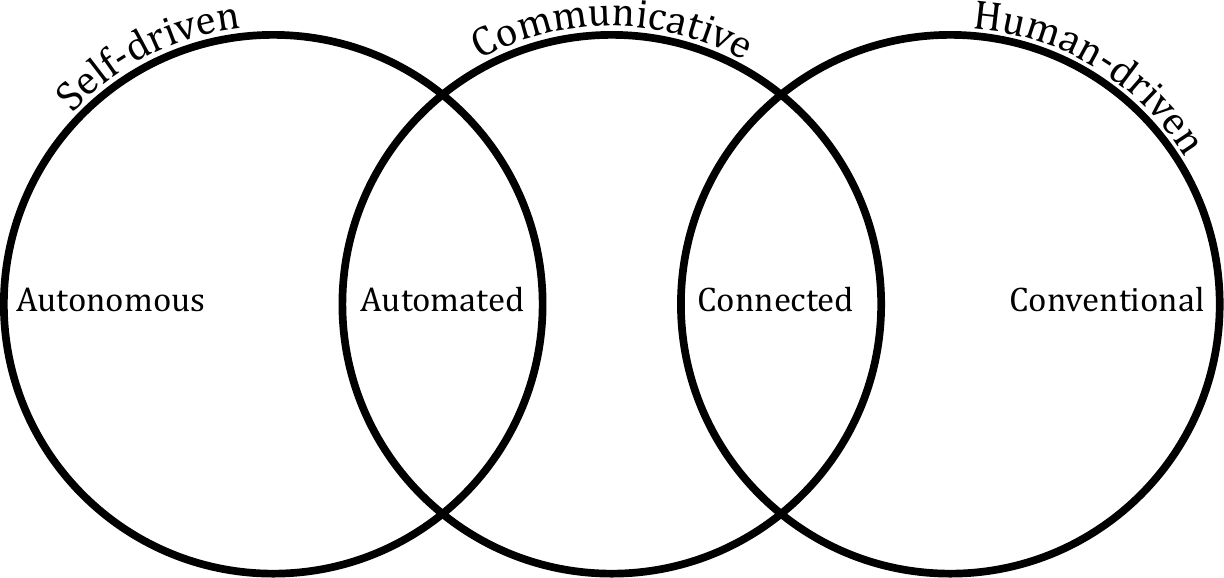}
\caption{Vehicles\textsc{\char13} classification based on their functionality.}\label{fig:Vann-CAV}
\end{figure}

The term “autonomous” or “self-driving” refers to the class of vehicles capable of performing the driving task without human intervention. While an autonomous vehicle is only concerned with operating itself, an Automated Vehicle (AV) also communicates with other vehicles, infrastructure, or cloud to enhance the transportation system\textsc{\char13}s performance. A connected vehicle (CV) also exchanges information, but a human driver controls the vehicle. Fig.~\ref{fig:Vann-CAV} distinguishes these classes of vehicles based on their general functionality and level of autonomy.

\cite{NHTSA2016} recently adopted the SAE International definitions for six levels of automation ranging from SAE level 0, being a conventional vehicle, to SAE level 5, being a fully autonomous vehicle. Intermediate levels share responsibilities between the driver and the automation system.  In this study, the traffic stream consists of conventional vehicles as well as communicative vehicles\textemdash those which comply with SAE levels 0 to 5 and capable of receiving instructions on their movement.

Recent attention has focused on designing signal control algorithms to incorporate advanced vehicle technologies into the transportation system. In one of the first studies that consider intersections, \cite{Dresner2004} set up an agent-based algorithm to accept incoming requests from AVs to reserve a set of tiles to complete their maneuver within the intersection\textsc{\char13}s conflict zone. Their algorithm assigns the right-of-way by granting the reservations with no conflict based on first-come-first-serve order in each lane.  The research assumes a fully automated environment with no conventional signal indication. Several other studies have also explored different options to prioritize vehicles\textsc{\char13} crossing sequence under a variety of settings. (\cite{Ahmane2013,Elhenawy2015,Feng2015,He2012,Levin2016,Li2006,Li2013,Tachet2016,Wu2013,Xie2012,Yan2008,Zhong2015}).

\cite{Li2014} proposed a rolling-horizon procedure to make joint decisions on vehicles\textsc{\char13} trajectory and signalization at an isolated intersection with AVs only and with two conflicting through movements. Their algorithm uses the arrival information at a detection distance to compute a four-component trajectory for every vehicle. The algorithm enumerates a set of binary signal decisions and selects the trajectories with the lowest average travel time delay. Several studies proposed algorithms to improve on adaptive signal control logic using the high-resolution data brought by automated/connected vehicle technology (\cite{Lee2012,Lee2013,Goodall2013,Feng2016,Yang2016,LeVine2016}). Beyond isolated intersections, \cite{He2015} proposed a real-time speed advisory algorithm that reduces the fuel consumption of a connected vehicle traveling along an arterial with coordinated signals. \cite{Priemer2009} developed a signal phase control algorithm that reduces the queue length at a signal network using V2I communication.

This study develops a trajectory-based optimization algorithm applicable to an isolated intersection under mixed-traffic.  The algorithm optimizes both the signal control and the trajectories of AVs assuming there is no lane changing past the point of initial communication of the vehicle and the intersection controller. The algorithm assumes the traffic is undersaturated, i.e., there is no need to track vehicles over more than one cycle.  We develop a recursive model to establish the follower-lead dependency among vehicles in each incoming lane. Depending on the type and position of each vehicle, we solve either a trajectory optimization (for automated vehicles) or trajectory estimation (for conventional vehicles) to obtain the respective trajectory. Due to the pivotal role of an AV lead trajectory on the movement of its followers, we formulate and solve a non-convex optimization problem that minimizes the travel time delay of vehicles within the available detection range. The control algorithm also adjusts the signal control settings.

The following section reviews the literature on AVs and intersection management.  The third section provides an overview of the proposed methodology, while the fourth section presents numerical results of the simulation experiments implementing our new approach, along with comparisons to conventional actuated control simulated in VISSIM. The final section discusses the performance of the algorithm, proposes potential improvements, and poses questions for future research.

\section{Literature Review}
\label{S:2}

This section reviews published studies on integrating AVs with intersection operations. Depending on the control logic, we classify the literature into two broad groups: (1) reservation-based, i.e., models that allocate the right of way of the intersection based on a set of pre-defined rules (2) trajectory-based, i.e., models that jointly optimize AV trajectories and signal control.

Under the first category, \cite{Dresner2004} developed the Autonomous Intersection Management (AIM) system, an algorithm that either rejects or grants requests of automated vehicles to reserve a block of space-time in the intersection. They designed AIM to replace traffic lights and stop signs. The paper indicates that the AIM system can coordinate vehicles\textsc{\char13} crossing sequence more efficiently than conventional traffic lights. In subsequent publications, the AIM was extended to: make  the control policy compatible with  communicative semi-autonomous vehicles, (\cite{Au2015,Dresner2006a,Dresner2006}); manage priorities for side traffic under unbalanced demand, (\cite{Au2011}); improve  estimation of  vehicle arrivals to the stop bar, (\cite{Au2010}); and coordinate  a network of multiple interconnected intersections, (\cite{Hausknecht2011}). Several other studies also supported intersection control logics without the presence of the traditional signal  (\cite{Elhenawy2015,Tachet2016,Wu2013,Yan2009}). These types of algorithms may have limitations on minimizing delay because of: (1) frequent switches of right-of-way may disrupt platoons resulting in higher total travel time for the intersection; (2) eliminating traffic signal heads makes the system less expectable for conventional vehicles; (3) they do not consider the additional benefit stemming from optimizing AV trajectories.

\cite{Feng2015} adopted a broader perspective of reservation-based models after they formulated and solved a bi-level problem that feeds the connected vehicle data into an adaptive signal control logic. Their algorithm predicts spatial information of conventional vehicles using the connected vehicles\textsc{\char13} real-time data. Compared to a fully-actuated control simulated in VISSIM, the proposed signalization reduced total delay by 16\% under higher rates of CVs. Despite addressing responsiveness to conventional vehicles, their approach limits decision-space to determining priorities within the conflict zone of an intersection.

Under the second category, trajectory-based optimization algorithms expand the decision space to improve intersection performance. \cite{Li2014} proposed a rolling-horizon process to solve a trajectory-based optimization for an intersection with two conflicting through movements assuming only AV traffic. The algorithm estimates the trajectory of an AV as a function of its speed and position in the traffic stream. An upper level control algorithm enumerates all the feasible switches between two conflicting movements within a certain time interval.  Their algorithm implements the combination of signal and trajectory decisions that yield the least average delay.  However, the signal control problem was solved supposing only two straight movements which makes it feasible to enumerate all timing options. Moreover, their methodology does not take into account presence of conventional vehicles with no communication capabilities.

\cite{Guler2014} devised and simulated an algorithm to enumerate sequences to serve connected vehicles at an intersection with two through movements. Promoting platoons, they associated higher delay reduction with higher penetration rate of connected vehicles in traffic. Recently, \cite{Yang2016} also evolved this work of \cite{Guler2014} to solve a trajectory-based optimization for various types of AV/CV approaching an isolated intersection. Once a vehicle arrived at a communication range from center of intersection, an upper level algorithm provided departure sequence of vehicles through a branch and bond tree search. Next, the lower level algorithm obtains the set of piecewise-linear trajectories to obtain the least amount of average delay and number of stops. Through simulation experiments, the method improved on actuated signal control with at least 50\% penetration rate of automated and connected vehicles in traffic stream. Besides considering an over-simplified intersection with two through movements, a piece-wise linear trajectory form fails to address the deceleration/acceleration behavior of vehicles which can lead to underestimating overall delay.

In summary, there are two types of algorithms that mainly differ in the extend of control decisions, i.e. signal scheduling and trajectory planning. The family of reservation-based algorithms mostly focuses on obtaining the optimal sequence of serving lanes by sorting incoming requests from AVs. Trajectory-based algorithms, however, further benefit the connectivity and programmability of AVs to prepare them for the optimal departure far ahead from the stop bar. On the other hand, considering the real-time execution of the algorithms, the additional complexity in trajectory-based models could necessitate more computational resources to be operational.


This study considers a mixed vehicle environment, aims to enhance adaptive signal control, and takes advantage of AV presence to adjust their trajectories and further minimize the overall travel time.

\section{Methodology Overview}
\label{S:3}
\begin{figure}[h]
\centering
\includegraphics[width=1\linewidth]{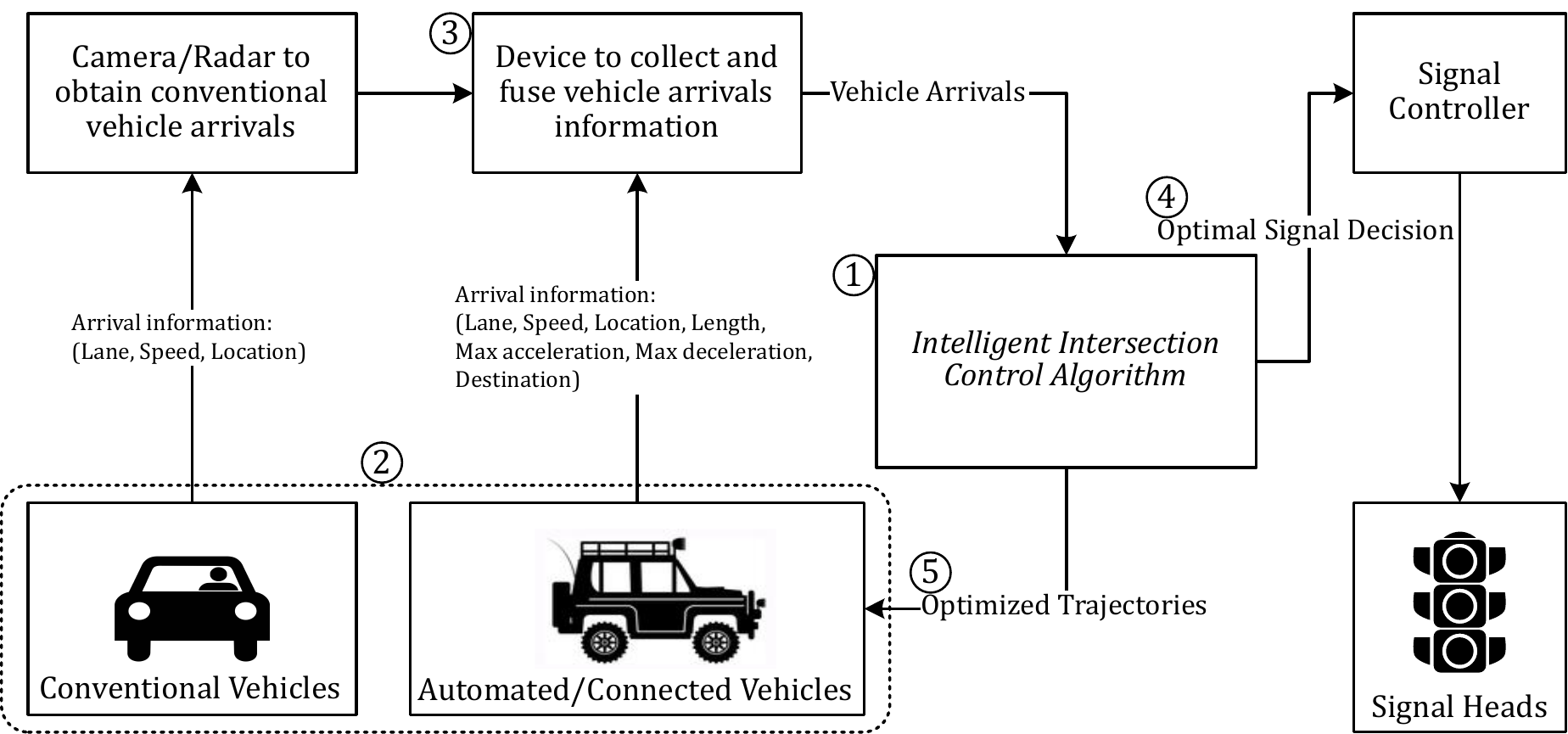}
\caption{Intelligent Intersection Control System (IICS).}\label{fig:FlowCharts}
\end{figure}

In this section, we develop models and sub-models to optimize intersection performance using arrival information of AVs and conventional vehicles. We devise an algorithm to process the arrival data at a lane-specific detection range on a real-time basis. The traffic composition of (AVs, with two-way V2I connectivity, and conventional vehicles, with no ability to communicate) is an input to the algorithm. The algorithm aims to minimize travel time delay of arriving AVs by preventing unnecessary stops within the communication range. This goal is achievable by synchronizing the signal control parameters only for undersaturated condition as queues are inevitable beyond that. Fig.~\ref{fig:FlowCharts} schematically shows the proposed Intelligent Intersection Control System (IICS).

Initially, the central computer, marked as 1 in Fig.~\ref{fig:FlowCharts}, receives each vehicle\textsc{\char13}s arrival information, marked as 2 and 3, from a detection distance\textemdash the center line distance from detection range to the stop bar\textemdash in each lane. Next, the algorithm computes the parameters associated with the movement of vehicles toward the intersection. The proposed trajectory model minimizes the travel time delay of the automated and estimates the movement of the conventional vehicles. The algorithm implements a modified adaptive control logic based on computed trajectories, marked as 4. Finally, the algorithm provides two sets of outputs: (1) the signalization pattern, marked as 4; (2) the trajectory of each AV to optimize travel time from the detection location to the stop bar, marked as 5. The procedure is repeated each time a new vehicle arrives.

The next subsections present the proposed model to describe the movement of detected vehicles. In this section, we model the movement of approaching vehicles. The trajectory computation\textsc{\char13}s state equation involves sub-models to predict trajectory of a conventional vehicle or optimize trajectory of an automated vehicle. The adaptive signal control logic is described in section.~\ref{S:3.2}.


\subsection{Algorithms for Trajectory Optimization/Estimation}
\label{S:3.1}
The vehicle type, i.e. conventional or automated, and its position on the lane determine the trajectory computation process. We define a lead vehicle as a vehicle that moves free from influence of others. Any vehicle\textsc{\char13}s movement can affect its follower depending on a variety of factors. Considering the cumulative effect in a lane, the lead vehicles have a key role in intersection delay. Finally, we will formulate and solve a specific optimization problem to minimize the travel time delay of an automated lead vehicle. We formulate an equation to address the follower\textendash lead dependencies among a set of vehicles in each lane. The Automated vehicle Trajectory Optimization (ATO) reflects this effect by recursion. A solution to ATO model guarantees that trajectory of all the $k-1$ first vehicles can affect the trajectory of $k$th vehicle.

Data:
\begin{enumerate}[leftmargin=1cm, labelsep=0.1cm, align=left, itemsep=-0.1cm, font=\small]
\item[$\mathcal{L}$] the set of incoming lanes, $l \in \mathcal{L}$
\item[$\mathcal{K}_l$] the ordered set of vehicles in lane $l$, $\forall \ l \in \mathcal{L}$
\item[$type_{kl}$] the type of vehicle $k$ on lane $l$, $type_{kl} \in \{AV: automated/connected \ vehicle, CV: conventional \ vehicle\},\quad\forall \ l \ \in \mathcal{L},\quad\forall \ k \ \in \mathcal{K}_l$
\item[$spt_{kl}$] spatial information of vehicle $k$ on lane $l$, including detection time stamp $t_0$, location $d_{kl} (t_0 )$, speed $V_{kl} (t_0 )$, and movement of AVs $m\in M=\{left,\ straight,\ right\}$
\item[$att_{kl}$] set of attributes for vehicle $k$ on lane $l$ including maximum acceleration rate $a_{kl}^{max+}$, maximum deceleration rate $a_{kl}^{max-}$, vehicle longitudinal length $L_{kl}$, desired speed $V_{kl}^{des}$
\item[$spd_{m}$] set of speed limits $V_m^{max}$ and $V_m^{cross}$ within the detection range and at the stop bar for movement $m \in M$
\item[$\Phi$] set of active phases to serve all available movements at the intersection, $\phi \in \Phi$
\item[$\Delta$] the phase-lane incident matrix $\Delta=[\delta_{\phi l} \ \forall \ l \in \mathcal{L},\quad\forall \ \phi \in \Phi]$, where $\delta_{\phi l}$ is 1 if lane $l$ belongs to phase $\phi$, 0 otherwise
\item[$sig$] the array that contains signal control events, $sig_{\phi}=(t_{s_{\phi}},G_{\phi},Y_{\phi},{AR}_{\phi} )$

	\begin{enumerate}[leftmargin=2cm, labelsep=0.1cm, align=left, itemsep=-0.1cm, 		font=\small]
    	\item[$t_{s_{\phi}}$] the time when the green interval for phase $\phi$ begins
        \item[$G_{\phi}$] the duration of green interval for phase $\phi$
        \item[$Y_{\phi}$] the duration of yellow interval for phase $\phi$
        \item[${AR}_{\phi}$] the duration of all-red interval for phase $\phi$
	\end{enumerate}
\end{enumerate}

Variable:
\begin{enumerate}[leftmargin=1cm, labelsep=0.1cm, align=left, itemsep=-0.1cm, font=\small]
\item[$traj_{kl}$] to be the trajectory of vehicle $k$ on lane $l$, $\{d_{kl}(t): t \in T_{kl} \}\quad\forall \ l \in \mathcal{L},\quad\forall \ k \in \mathcal{K}_l$
\begin{enumerate}[leftmargin=2cm, labelsep=0.1cm, align=left, itemsep=-0.1cm, font=\small]
    	\item[$T_{kl}$] To be the time interval from vehicle detection to its departure at the stop bar
        \item[$d_{kl}(t)$] to be the center-lane distance of vehicle $k$ on lane $l$ to the stop bar
	\end{enumerate}
\end{enumerate}

Recursive State Equation for Automated vehicle Trajectory Optimization (ATO):
\begin{equation} \label{eq:ATO}
   traj_{kl}=  \left\{
                \begin{array}{ll}
                  FTO(sig, spt_{kl}, att_{kl}, traj_{(k-1)l})   \quad \textnormal{for} \quad k \in K_l \backslash \{1\}, \ type_{kl}=AV\\
                  FTE(spt_{kl}, att_{kl}, traj_{(k-1)l})   \quad \textnormal{for} \quad k\in K_l\backslash \{1\}, \ type_{kl}=CV\\
                  LTO(sig, spt_{kl}, att_{kl}, spd_m)   \quad \textnormal{for} \quad k \in \{1\}, \ m \in M  , type_{kl}=AV\\
                \end{array} \ \forall \ l \in \mathcal{L}
              \right.
\end{equation}

Sub-models to ATO model:
\begin{enumerate}[leftmargin=1cm, labelsep=0.1cm, align=left, itemsep=-0.1cm, font=\small]
\item[$FTO(.)$] to be Follower vehicles Trajectory Optimizer for AVs
\item[$FTE(.)$] to be Follower vehicle Trajectory Estimator for CVs
\item[$LTO(.)$] to be Lead vehicle Trajectory Optimizer for AVs
\end{enumerate}

Eq.~\ref{eq:ATO} represents the trajectory of $k$th vehicle in lane $l$ to be a function of signalization, characteristics of the vehicle itself, speed limits, and recursively trajectory of its lead vehicle. However, in absence of dependency to trajectory of vehicle in front, as it will be discussed in section.~\ref{S:3.1.2}, the model considers the vehicle as a lead vehicle by itself. The sub-models (i.e. LTO, FTO, and FTE) to ATO problem will be discussed next.

\subsubsection{Lead vehicle Trajectory Optimization (LTO)}
\label{S:3.1.1}
In this section, we formulate and solve the Lead vehicle Trajectory Optimization (LTO) problem. The problem aims to minimize an automated lead vehicle\textsc{\char13}s travel time delay. An appropriate functional form for automated vehicle trajectory\textemdash also called the space-time function\textemdash should meet several qualifications: (1) to offer sufficient flexibility to capture and improve the movement of AVs; (2) to produce an accurate trajectory that the vehicle can implement; (3) to belong to a family of functions uncomplicated to be parametrized.

Here we formulate the LTO problem assuming the vehicle travels the detection range in three ordered stages. The first and third stages adjust the speed by a constant acceleration or deceleration rate, while the vehicle maintains a constant speed during the second stage. The three-stage trajectories closely models the behavior of the lead AV, however, a follower AV may peruse a more general strategy. Depending on the arrival parameters, the LTO solution may omit any of three stages by assigning it zero time duration. Eqs.~\ref{eq:stage1} to ~\ref{eq:stage3} provide the formulas to compute the travel time of each stage using fundamental motion equations.

\begin{equation} \label{eq:stage1}
 \Delta t_{kl,1} = \frac{v_2-v_1}{a_1} \quad \forall \ l \in \mathcal{L},\quad\forall \ k \in \mathcal{K}_l
\end{equation}
\begin{equation} \label{eq:stage2}
 \Delta t_{kl,2} = (d_0-\frac{v_2^2-v_0^2}{2 a_1}-\frac{v_3^2-v_2^2}{2 a_3})/v_2 \quad \forall \ l \in \mathcal{L},\quad\forall \ k \in \mathcal{K}_l
\end{equation}
\begin{equation} \label{eq:stage3}
 \Delta t_{kl,3} = \frac{v_3-v_2}{a_3} \quad \forall \ l \in \mathcal{L},\quad\forall \ k \in \mathcal{K}_l
\end{equation}
\begin{equation} \label{eq:Tklsum}
 T_{kl} = \sum_{n=1}^{3} \Delta t_{kl,n} \quad \forall \ l \in \mathcal{L}, \quad \forall \ k \in \mathcal{K}_l
\end{equation}

Where, to avoid lengthy mathematical expressions, we excluded vehicle and lane indices and used a more compact notation as:
\begin{enumerate}[leftmargin=1cm, labelsep=0.1cm, align=left, itemsep=-0.1cm, font=\small]
\item[$d_0$] center-lane distance from the stop bar at detention moment, $d_{kl} (t_0 )$
\item[$v_0$] initial speed of vehicle at the detection, $V_{kl} (t_0 )$
\item[$v_2$] constant speed at second stage, $V_{kl} (t_0+dt) \ \forall \ dt \in [\Delta t_1,\Delta t_1+\Delta t_2 ]$
\item[$v_3$] discharge speed of vehicle at the stop bar, $V_{kl} (t_0+T_{kl} )$
\item[$a_1$] acceleration/deceleration rate at first stage
\item[$a_3$] acceleration/deceleration rate at third stage
\end{enumerate}

Assuming the ideal travel time associated with the $k$th vehicle in lane $l$ to be the time required to travel the detection range at its desired speed, we compute the total travel time delay as given in Eq. ~\ref{eq:delay}:

\begin{equation} \label{eq:delay}
 D_{kl}(T_{kl}) = T_{kl}-\frac{d_0}{V_{kl}^{des}} \quad \forall \ l \in \mathcal{L},\quad\forall \ k \in \mathcal{K}_l
\end{equation}
Where $D_{kl}$ denotes the total travel time delay of vehicle $k$ on lane $l$

Next, we present the LTO mathematical program by minimizing vehicle\textsc{\char13}s travel time delay subject to restrictions from signalization, vehicle attributes, and regulation, as follows:

\begin{align}
    &(LTO) \min_{v_2,v_3,a_1,a_3} \label{eq:LTO}
    \begin{aligned}[t]
       &D_{kl}(T_{kl})
    \end{aligned}  \\
    &\text{subject to} \notag \\
    & t_{s_{\phi}} \leq \delta_{\phi l} T_{kl} \leq t_{s_{\phi}}+G_{\phi}+Y_{\phi} \\
    & v_2 \leq V_m^{max} \\
    & v_3 \leq V_m^{cross} \\
    & a^{max-} \leq a_1 \leq a^{max+}\\
    & a^{max-} \leq a_3 \leq a^{max+}
\end{align}
Where:
\begin{enumerate}[leftmargin=1cm, labelsep=0.1cm, align=left, itemsep=-0.1cm, font=\small]
\item[$D_{kl}(T_{kl})$] Travel Time Delay of vehicle $k$ on lane $l$ as a function of its travel time $T_{kl}$
\item[$t_{s_{\phi}}$] time when phase $\phi$ starts
\item[$\delta_{\phi l}$] equal to one if lane $l$ belongs to phase $\phi$, 0 otherwise
\item[$G_{\phi}$] green duration for phase $\phi$
\item[$Y_{\phi}$] yellow duration for phase $\phi$
\item[$v_2$] constant speed at second stage
\item[$v_3$] discharge speed of vehicle at the stop bar
\item[$a_1$] acceleration/deceleration rate at first stage
\item[$a_3$] acceleration/deceleration rate at third stage
\item[$a^{max-}$] maximum deceleration rate
\item[$a^{max+}$] maximum acceleration rate
\item[$V_m^{max}$] maximum allowable speed within the detection range
\item[$V_m^{cross}$] maximum allowable speed at the stop bar
\end{enumerate}
The first set of bound\textendash constraints keeps only solutions that the travel time belongs to the green or yellow interval. The next constraint guarantees the speed profile of vehicle does not exceed the maximum allowable speed starting from the second stage. The third constraint controls the discharge speed depending on the type of movements i.e. left turn, straight through, or right turn. The last set of constraints limit any acceleration or deceleration rate to the range that vehicle can execute.

The non-positive-definite Hessian matrix of the travel time delay, as appears in both LTO objective function and set of constraints, causes non-convexity. Even at four decision variables, standard non-linear methods are unable to distinguish the global optimal from non-optimal or infeasible solutions. By taking advantage of the size and structure of LTO problem, we develop an exact algorithm to find the global minimum.

The gradient of the objective function reveals that the travel time delay monotonically varies with respect to acceleration or deceleration rate in the first and the last stages. The necessary condition\textemdash setting gradient to zero\textemdash also indicates absence of any stationary point. Thus, considering the non-convex feasible region, the optimal solution must belong to the boundaries of the feasible region. Therefore, we solve LTO, model.~\ref{eq:LTO}, by devising an algorithm to minimize the travel time delay on the boundaries and not inside of feasible region, as shown by Algorithm.~\ref{alg:LTO}.
\begin{algorithm}[h]
\caption{AV Lead vehicle Trajectory Optimizer} \label{alg:LTO}
\begin{algorithmic}[1]
\Require signal control events, vehicle arrival information, vehicle attributes, and speed limits
\Ensure valid trajectory with minimal delay for the AV

\Procedure{LTO\_Exact\_Solver}{$sig,spt_{kl},att_{kl},spd_{kl}$}
\State $D_{kl}^* \gets M$ \Comment{M to be a relatively large value}
\State $flag \gets 0$

\For{$counter$ = 1:4} \Comment{LTO has four variables: $v_2,v_3,a_1,a_3$}
	\State Select a new variable
    \State Set non-selected variables to limit(s) based on bound-constraints
    \State Obtain the bounds on the selected variable
    \State Solve the remaining single-variable constrained problem by minimizing travel time delay over corrected bound
    \State  $D_{kl} \gets$ travel time delay corresponding to selected and computed variables
    \If{$D_{kl} < D_{kl}^*$} \Comment{current solution can be improved}
    \State $D_{kl}^* \gets D_{kl}$
    \State $flag \gets 1$
    \EndIf
\EndFor
\If{$flag = 1$}

	\Return{$D_{kl}^*$} \Comment{optimal solution}
\Else

	\Return{LTO problem is infeasible}
\EndIf
\EndProcedure
\end{algorithmic}
\end{algorithm}

\subsubsection{Follower Vehicle Trajectory Optimization}
\label{S:3.1.2}
Similar to the car following model for conventional vehicles, the Follower vehicle Trajectory Optimization (FTO) model intends to compute the trajectory of an automated follower. If the vehicle in front is going to depart at time $t_{(k-1)l}^{dep}$, assuming that the saturation headway at the stop bar is $s_h$, an automated follower can depart no sooner than $t_{(k-1)l}^{dep}+s_h$. Therefore a hypothetical trajectory for vehicle $k$ can be computed by lagging trajectory of vehicle $k-1$ to the saturation headway $s_h$. The hypothetical trajectory may not be compatible to the arrival of the $k$th vehicle.

Equivalently, the automated follower may fail to depart at the saturation headway if it is not close enough, or moves much slower, relative to its lead vehicle. Thus, even at the maximum acceleration rate, the automated vehicle may not be able to reduce its time headway down to the saturation time headway. Under this case, the proposed algorithm no longer considers that automated vehicle as a follower, and optimizes the trajectory as a lead vehicle by solving the LTO problem instead.

\begin{algorithm}[h]
\caption{AV Follower vehicle Trajectory Optimizer} \label{alg:FTO}
\begin{algorithmic}[1]
\Require trajectory of the lead vehicle, follower attributes, follower vehicle arrival information, and signal control events
\Ensure valid trajectory with minimal departure headway for automated follower

\Procedure{FTO\_Solver}{$sig,spt_{kl},att_{kl},traj_{(k-1)l}$}
\State $t_{kl}^{dep,hypo} \gets \max \{t_{s_{\phi}}, t_{(k-1)l}^{dep}+s_h \}$ \Comment{set the hypothetical earliest departure time to either initiation of green or departure at saturation headway}
\State{$dt \gets t_{kl}^{dep,hypo}-t_{(k-1)l}^{dep}$}
\State Construct the hypothetical trajectory, $traj_{kl}^{hypo}$ \Comment{by lagging all timestamps of $traj_{(k-1)l}$ up to $dt$}
\State $flag \gets 0$

\For{$index \in $trajectory point indices on $traj_{kl}^{hypo}$} \Comment{Searching for a feasible transition from arrival point to the hypothetical trajectory}
   \If{feasible deceleration/acceleration from detection speed to speed at $index$ exist} \Comment{Earliest hypothetical trajectory is feasible }
    \State {construct the earliest trajectory, $traj_{kl}$, by the feasible transition appended by all trajectory points of $t_{kl}^{dep,hypo}$ with the timestamps greater than $index$ }
    \State $flag \gets 1$
    \EndIf
\EndFor
\If{$flag = 1$}

	\Return{computed $traj_{kl}$}
\Else

	\Return{$traj_{kl}=LTO\_Exact\_Solver(sig,spt_{kl},att_{kl},spd_{kl})$}\Comment{use algorithm.~\ref{alg:LTO} as vehicle $kl$ could not be a follower}
\EndIf
\EndProcedure
\end{algorithmic}
\end{algorithm}

The procedure within Algorithm.~\ref{alg:FTO} outputs the trajectory with minimum travel time that keeps headway higher than, or equal to, the saturation headway. The principal reason comes from the fact that the trajectory is constructed either by joining the transition component with the hypothetical trajectory, or by solving LTO problem. Under the former case, the trajectory has a headway greater than or equal to the saturation headway; the higher headway corresponds to the transition part while the equality associates with the hypothetical part. The latter case, when the LTO problem is solved, no trajectory point can have a headway lower than the saturation headway. Having a lower headway would be in contradiction to not finding any transition to the hypothetical trajectory at lines 6-11. Finally, the trajectory takes the minimum travel time resulting in the minimum achievable headway at the stop bar.

\subsubsection{Trajectory Estimation for Conventional Vehicles}
\label{S:3.1.3}
For undersaturated condition, we assume that a lead conventional vehicle would desire to maintain its speed as recorded when entering the detection range. The rest of this section explains the model to estimate a follower conventional vehicle\textsc{\char13}s trajectory.

Sensors on the roadside, i.e. radar or camera, obtain the lane, location and speed of conventional vehicles once they enter the detection range. Although they are non-communicative vehicles, conventional vehicles affect both signalization and trajectory computation for automated vehicles. Therefore, the Automated vehicle Trajectory Optimization (ATO) model predicts their movements through a Car-Following model. For the purpose of this study, we implement the \cite{Gipps1981} car-following model\textemdash as the FTE sub-model in Eq.~\ref{eq:ATO} \textemdash to estimate the trajectory of a conventional follower. Below is the equation that yields the speed profile of conventional follower vehicle:

\begin{align}\label{eq:Gipps}
 v_{kl}(t+\Delta t) = \min \{ v_{kl}(t)+2.5 a_{kl}^{max+} \Delta t (1-\frac{v_{kl}(t)}{v_{kl}^{des}}) \sqrt{0.025+\frac{v_{kl}(t)}{v_{kl}^{des}}}, \\ \notag
 a_{kl}^{max-} \Delta t+ \sqrt{a_{kl}^{max-}(2(d_{(k-1)l}(t)+L_{kl}-d_{kl}(t))+\Delta t (a_{kl}^{max-} \Delta t+v_{kl}(t))+\frac{v_{(k-1)l}(t)^2}{a_{kl}^{max-}}}
 \}
\end{align}
where:
\begin{enumerate}[leftmargin=1cm, labelsep=0.1cm, align=left, itemsep=-0.1cm, font=\small]
\item[$\Delta t$] time steps to compute trajectory points
\item[$v_{kl}(t)$] speed of follower vehicle $\Delta t$ unit of time after $t$
\item[$v_{(k-1)l}(t)$] speed of lead vehicle $\Delta t$ unit of time after $t$

\end{enumerate}

In the context of the model, Eq.~\ref{eq:ATO}, Algorithm.~\ref{alg:FTO} implements the Gipps Car-Following model to compute a conventional vehicle\textsc{\char13}s trajectory.

\begin{algorithm}[h]
\caption{Conventional Follower vehicle Trajectory Estimator} \label{alg:FTE}
\begin{algorithmic}[1]
\Require trajectory of lead vehicle, lead and follower\textsc{\char13}s attributes, follower vehicle arrival information
\Ensure trajectory of conventional follower

\Procedure{FTE\_Solver}{$spt_{kl},att_{kl},att_{(k-1)l},traj_{(k-1)l}$}
\State $t \gets $detection  time
\State $d \gets $follower initial distance to stop bar
\While{$d > 0$}
	\State{$t \gets t+\Delta t$}
    \State{Compute $v_{kl}(t)$ using Eq.~\ref{eq:Gipps}}
    \State{$a_{kl}(\tau) \gets \{\frac{v_{kl}(t+\Delta t)-v_{kl}(t)}{\Delta t} | \tau \in [t,t+\Delta t] \}$}
    \State{$d \gets \frac{1}{2} a_{kl}(\tau) \Delta t^2+v_{kl}(t)\Delta t$}
\EndWhile

\Return{computed pairs of $(t,d)$}
\EndProcedure
\end{algorithmic}
\end{algorithm}

\subsection{Adaptive Signal Control with Trajectory Optimization}
\label{S:3.2}
\begin{figure}[h]
\centering
\includegraphics[width=1\linewidth]{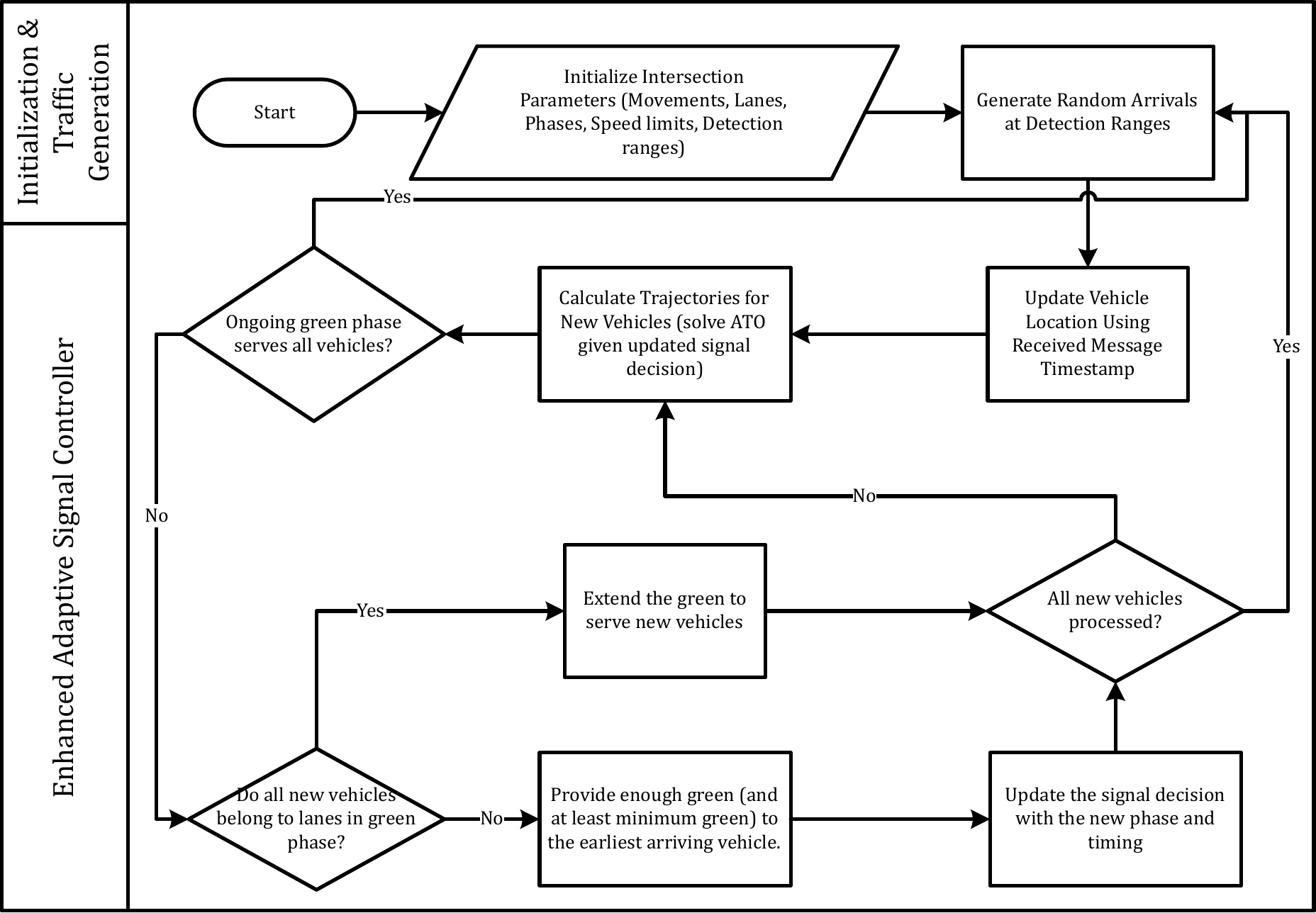}
\caption{Enhanced Adaptive Signal Control to use Trajectory Data.}\label{fig:EnhancedAdaptive}
\end{figure}

We devise an enhanced adaptive signal control logic based on trajectory information to make decisions on whether to extend or switch a phase. The proposed algorithm, as shown in Fig.~\ref{fig:EnhancedAdaptive}, re-evaluates the signal control status every time a new vehicle arrives. If the current signal phase can serve the earliest newly detected vehicle, it asks the signal controller for a green extension. Otherwise, it terminates the ongoing green and switches the right-of-way to the phase that leads to least travel time delay for incoming vehicles. The information from optimized trajectories allow for exact scheduling of phases. Although the algorithm has no control over conventional vehicles, they can have greater time headway as AVs travel the detection range in least time.

The timing for each phase should meet several practical criteria. Any green interval lower than a minimum green or higher than a maximum green causes too frequent or late phase switches which forces excessive delay to vehicles.

\begin{figure}[h]
\centering
\includegraphics[width=1\linewidth]{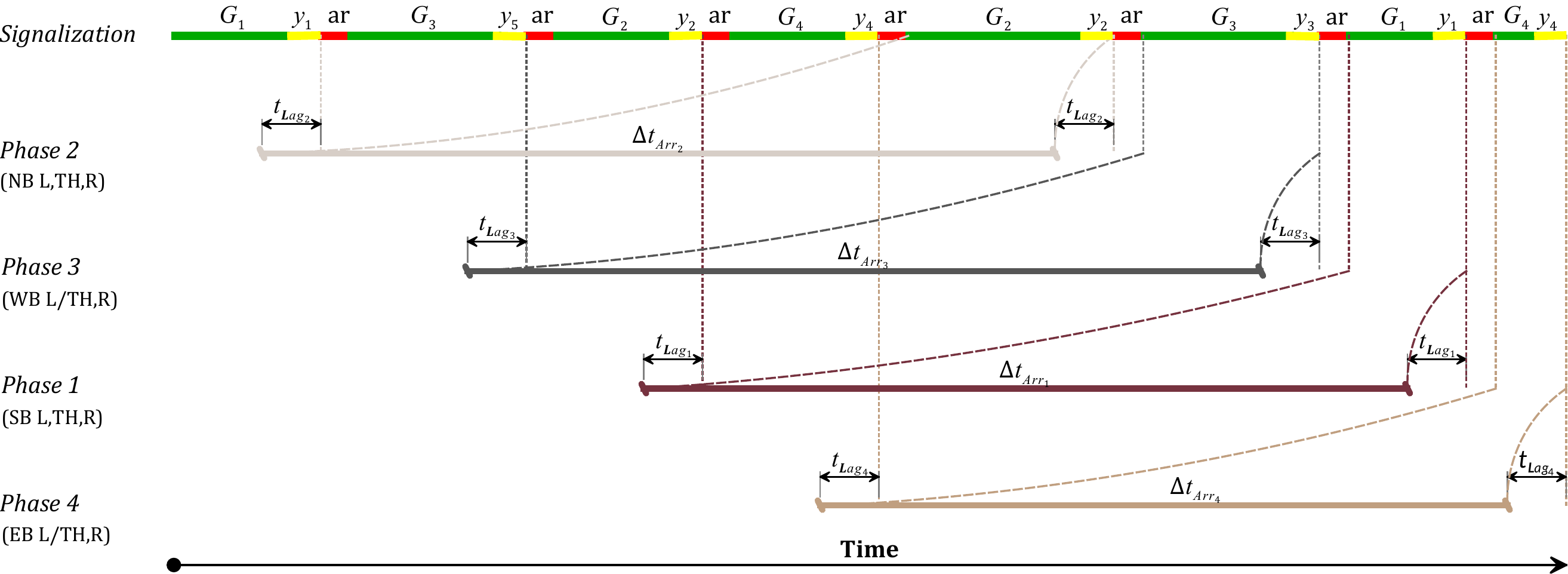}
\caption{Schematic Signal Control Plan($G_i$ and $y_i$ are the green and yellow duration for phase $i$; $\Delta Arr(i)$ indicates the arrival interval in phase $i$; $t_{Lag(i)}$ denotes the time before the end of yellow interval in phase $i$).}\label{fig:Schematic_signal_control}
\end{figure}

While the enhanced adaptive control logic mimics the traditional adaptive signal control strategy, the proposed algorithm makes decisions using the trajectories of AVs instead of calls coming from loop detectors or cameras. Fig.~\ref{fig:Schematic_signal_control} illustrates the association of each arrival interval and the corresponding signal decision for a few consecutive phases. The minimum time to travel the detection range necessitates a lag time, $t_{Lag}$, between the end of the green and the end of the associated arrival interval. In other words, the lag time represents the time between when the algorithm makes signal decision and when the corresponding traffic will depart at the stop bar.

\section{Algorithm Implementation and Numerical Results}
\label{S:4}
\begin{figure}[h]
\centering
\includegraphics[width=1\linewidth]{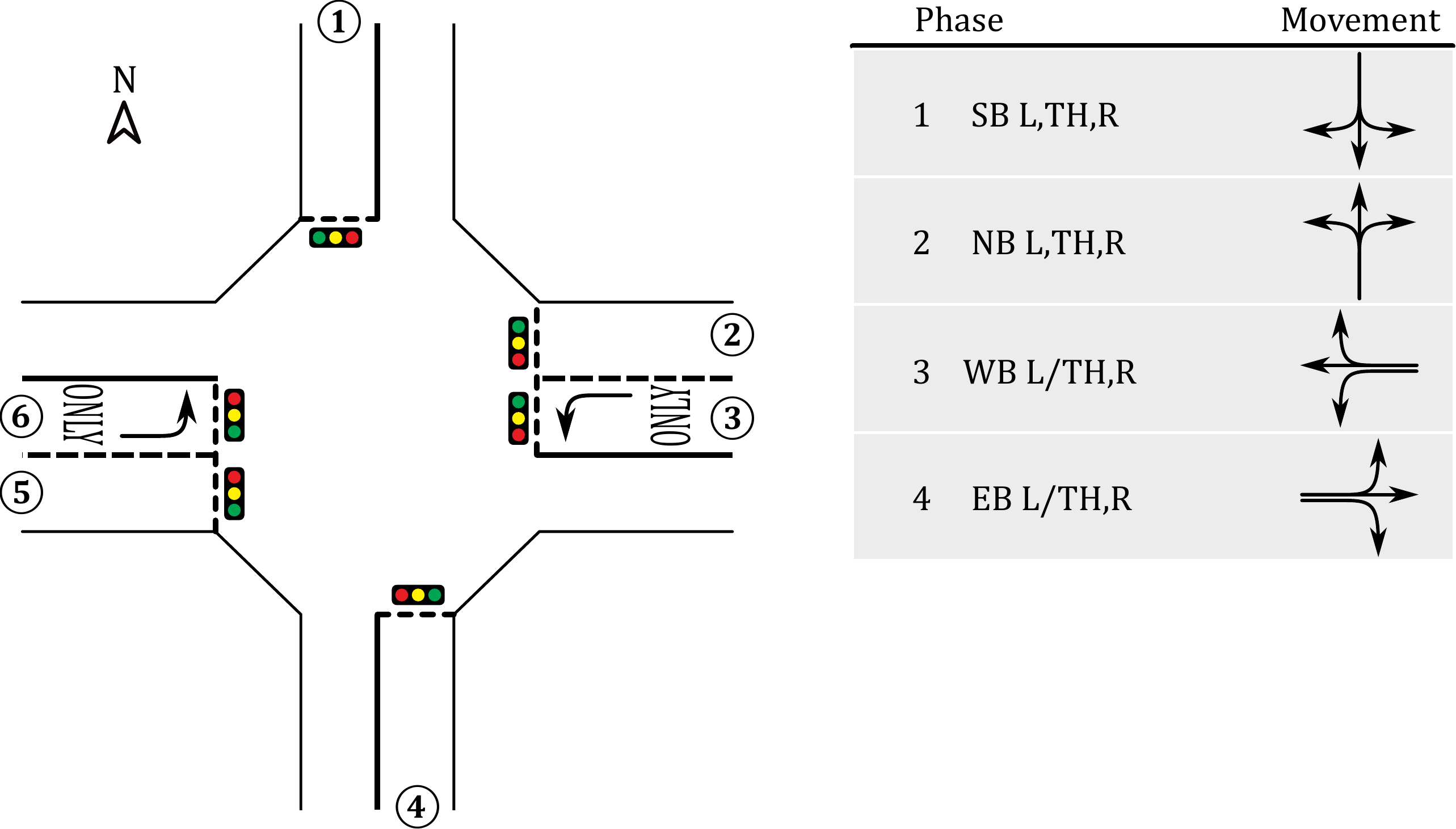}
\caption{A four-leg test intersection with six incoming lanes and four phases.}\label{fig:Intersection}
\end{figure}

We program the proposed IICS process in \cite{MATLAB} customized for a four-leg intersection with six incoming lanes as shown in Fig.~\ref{fig:Intersection}.

According to Fig.~\ref{fig:Schematic_signal_control}, the intersection includes: six approaching lanes, four departing lanes, and two approaches with exclusive left turn lanes.

We implement the IICS based on the following assumptions:

\begin{itemize}[leftmargin=1cm, labelsep=0.1cm, align=left, itemsep=-0.1cm, font=\small]
\item Sum of critical flows results in undersaturated conditions.
\item \cite{Gipps1981} car-following model predicts the conventional followers\textsc{\char13} movement, as described in section.~\ref{S:3.1.3}.
\item For simulation purposes, no data loss due to communication malfunction occurs.
\item Intersection\textsc{\char13}s geometry:
	\begin{itemize}
	\item The four-leg test intersection is on level terrain with six incoming lanes and all turning movements available, see Fig.~\ref{fig:Schematic_signal_control}.
	\end{itemize}
\item Operating conditions:
	\begin{itemize}
	\item Maximum allowable speed is 40 mph.
    \item Turning movements cross the stop bar at speeds no more than 30 mph.
    \item Minimum green time for all phases is 4.5 seconds.
    \item Yellow and all-red time are 1.5 seconds per phase.
    \item No lane changing occurs once a vehicle arrival is detected.
    \item No pedestrians are present in the vicinity of the intersection; therefore, no pedestrian phases are required.
	\end{itemize}
\item Traffic generation:
	\begin{itemize}
	\item Vehicles arrive at the communication distance and are simulated in a 15-minute period.
    \item The initial speeds of vehicles follow the triangular distribution with minimum, peak, and maximum values equal to 34 mph, 40 mph, and 44 mph (as 0.85, 1, and 1.1 factors of the maximum allowable speed), respectively.
    \item Vehicles, regardless of being automated or conventional, are capable of deceleration/acceleration of -15 ft/s$^2$ to 10 ft/s$^2$.
    \item Conventional vehicles\textsc{\char13} desired speed is 40 mph.
    \item The IICS algorithm uses a safe speed recommendation inside the communication area, however, in some cases the initial speed of individual vehicles might be slightly higher\textemdash as reflected by the triangular distribution with maximum of 44 mph.
	\end{itemize}
\end{itemize}

\subsection{Simulation Experiments}
\label{S:4.1}
This section reports the results of 3000 simulation experiments. Using the algorithm described in section.~\ref{S:3}, the following  four variables are used to formulate the test scenarios:

\begin{itemize}[leftmargin=1cm, labelsep=0.1cm, align=left, itemsep=-0.1cm, font=\small]
\item The detection range for all lanes varies from 500 to 3000 feet (ten equidistant values).
\item The AV ratio for all lanes varies from 0.3 to 1 (ten equidistant values); 1 being the traffic with all automated vehicles and no conventional vehicles.
\item Vehicle inter-arrival times follow the exponential distribution with rate parameter equal to the inverse of the average time headway. The average time headway for all lanes varies from 8 to 60 seconds (ten equidistant values).
\item The saturation headway at stop bars to be 1, 1.5, or 2 seconds. \cite{Lioris2017} showed that connectivity of vehicles can reduce the saturation headway by forming platoons.
\end{itemize}

\begin{figure}[h]
\centering
\includegraphics[width=.6\linewidth]{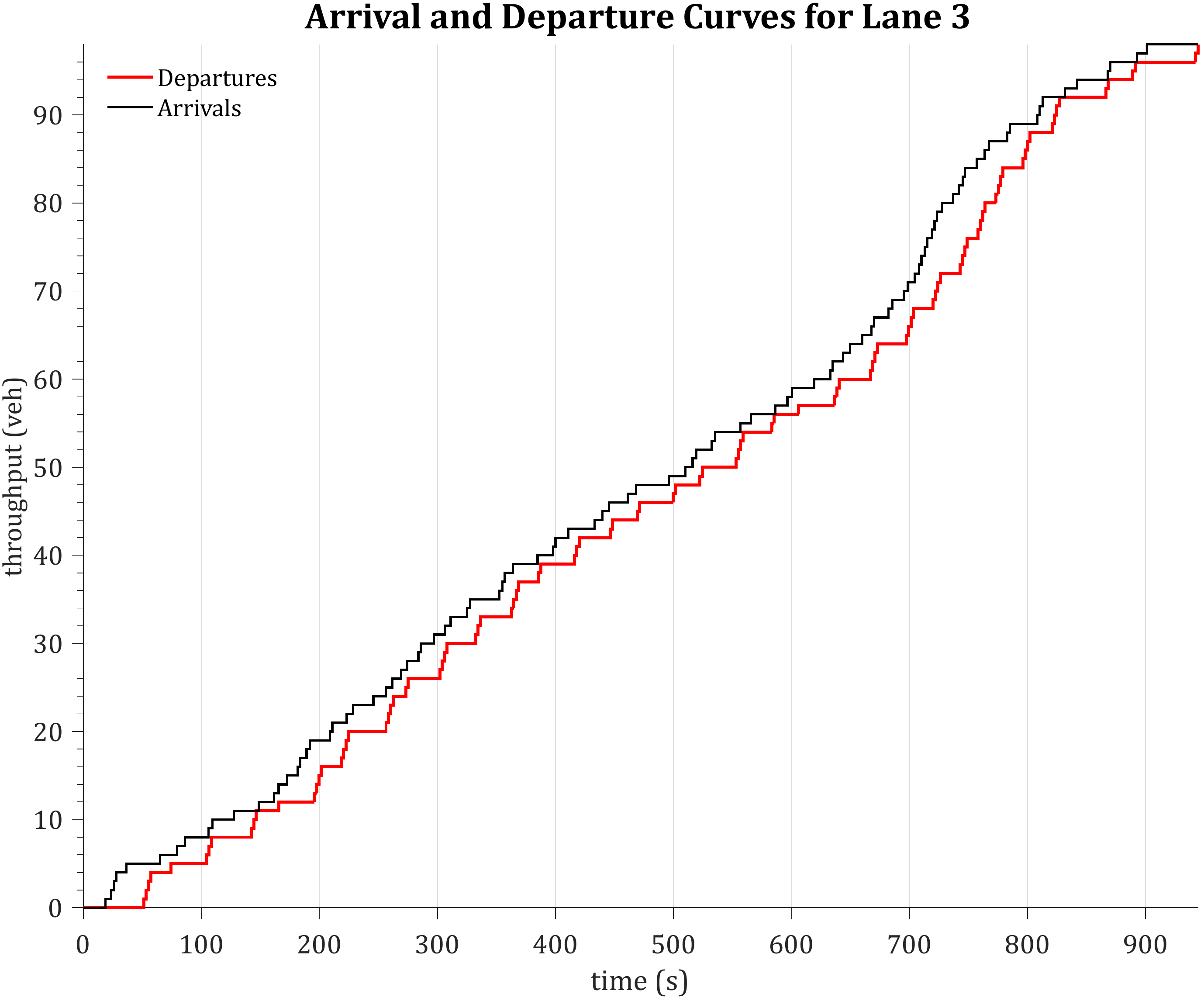}
\caption{Cumulative arrival and departure curves after 15 minutes of simulation (Notice departures occur at the stop bar while arrivals occur at the detection distance, in this case 1000 feet far from the stop bar.)}\label{fig:throughput}
\end{figure}

Fig.~\ref{fig:throughput} illustrates how the algorithm served arriving vehicles in a lane. The horizontal time gap between arrival and departure curve represents the time that each vehicle spent traveling through the detection prior to the stop bar.

\begin{figure}[h]
\centering
\includegraphics[width = .95\columnwidth]{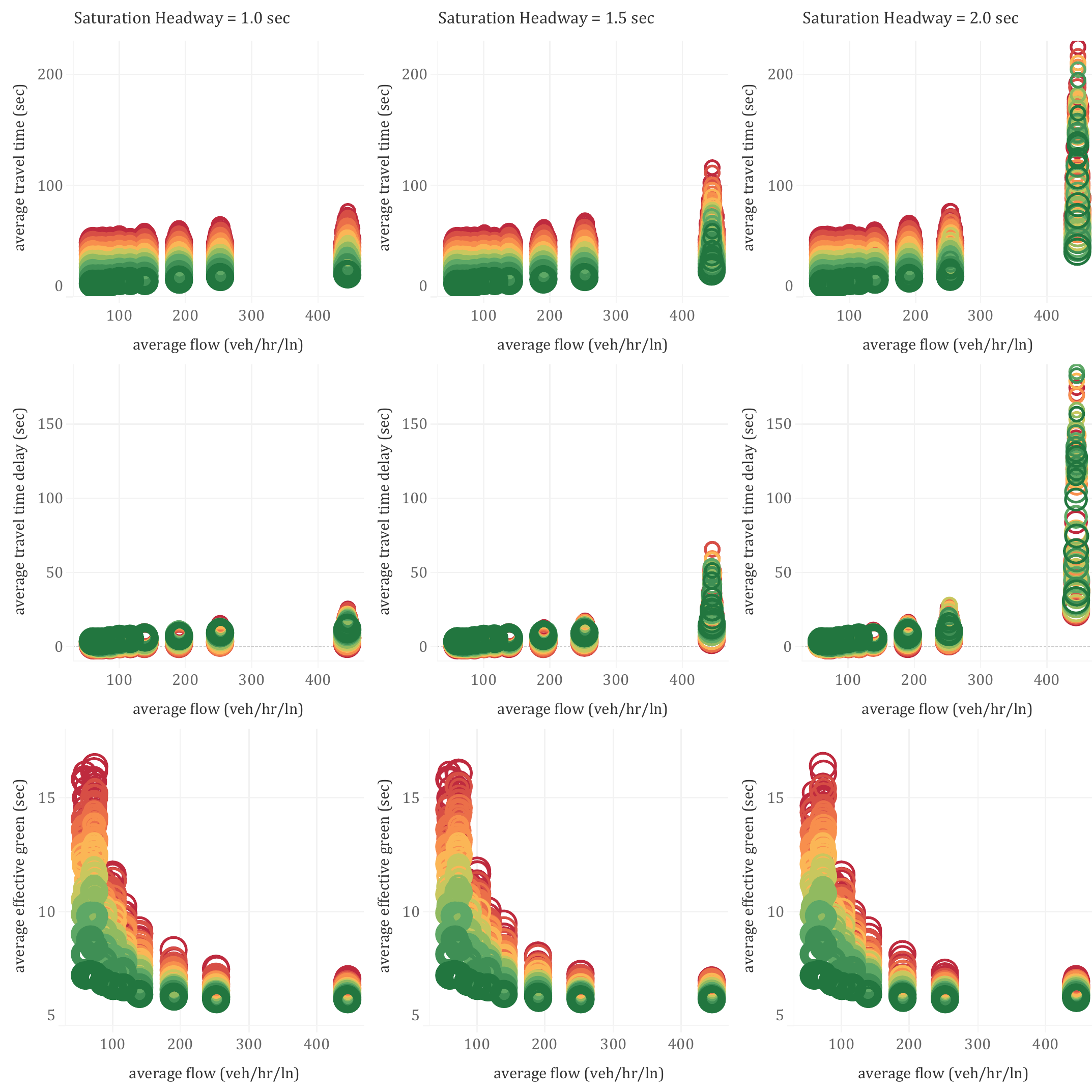}
\caption{Sensitivity analysis of results for 3000 scenarios (panels in same column associate with same saturation headway; panels in same row associate with same measure of effectiveness on the vertical axis; AV ratio: Automated/Connected Vehicle ratio in the traffic stream, the rest to be conventional vehicles.)} \label{fig:dataVis}
\end{figure}

In order to quantify the performance, a module monitors three outcome variables including: average travel time (measured from detection to departure at the stop bar), average delay (determined as the extra travel time relative to the free flow travel time), and average effective green (sum of actual green and yellow times). Under the set of assumptions stated in section.~\ref{S:4}, we observed the following patterns (Fig.~\ref{fig:dataVis}):

\begin{itemize}[leftmargin=1cm, labelsep=0.1cm, align=left, itemsep=-0.1cm, font=\small]
\item The average travel time and average delay surge as flow increases. For saturation headway equal to 2 second, once the sum of critical flows approaches the flow threshold, the operation becomes unstable\textemdash indicating queues are inevitable.
\item As expected, capacity increases as the saturation headway decreases (see the first row of panels in Fig.~\ref{fig:dataVis}).
\item The average delay slightly decreases as the detection range increases. The higher the available distance, the more flexibility in the trajectory optimization, which reduces the total delay.
\item The higher the ratio of AVs, the lower the average delay.
\item In higher flows, the IICS model assigns less amount of effective green to each phase (see the last row of panels in Fig.~\ref{fig:dataVis}). This reflects the increase in the likelihood that multiple phases would claim the right-of-way simultaneously as flow intensifies.
\item The IICS algorithm also allocates more green for higher communication distances to guarantee serving incoming vehicles. This trend is less noticeable for higher flow as the conflicting movements are more likely to prevent green extensions.
\end{itemize}

\subsection{Comparison to VISSIM}
\label{S:4.2}
In order to compare the proposed IICS framework with state-of-the-art intersection control, we model the same intersection under fully-actuated signal control in VISSIM. To keep the same basis for comparison, the 15-minute simulations in VISSIM: (1) have the same incoming flows and arrival distributions; (2) collect travel time information at distances equal to the detection range in IICS; (3) use the same saturation headways at the stop bar. The VISSIM implementation differs from the IICS model in terms of: (1) a fully actuated logic\textemdash using loop detectors for all incoming lanes; (2) traffic consists of only conventional vehicles.

\begin{figure}[h]
\centering
\includegraphics[width=.7\linewidth]{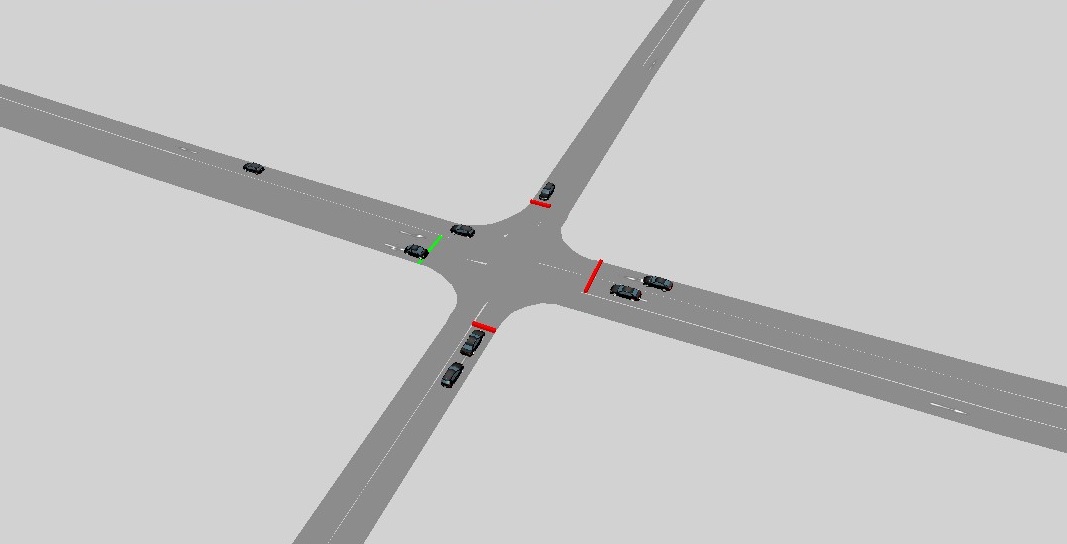}
\caption{Near undersaturated flow threshold, the implemented fully-actuated signal control VISSIM fails serving all conventional vehicles without queue formation.}\label{fig:VISSIM}
\end{figure}

For the calibration process, only a few parameters were adjusted. The default number of observed vehicles and the look ahead distance from the car-following model were increased. Those variations allow for a smoother simulation and a model closer to an automated environment where vehicles will gather more information from the surroundings than actual drivers, even though the actual model only simulates conventional vehicles. Fig.~\ref{fig:VISSIM} shows a snapshot of the fully actuated intersection control implemented in VISSIM.

\begin{figure}[h]
\begin{subfigure}{1\textwidth}
\centering
\includegraphics[width=1\linewidth]{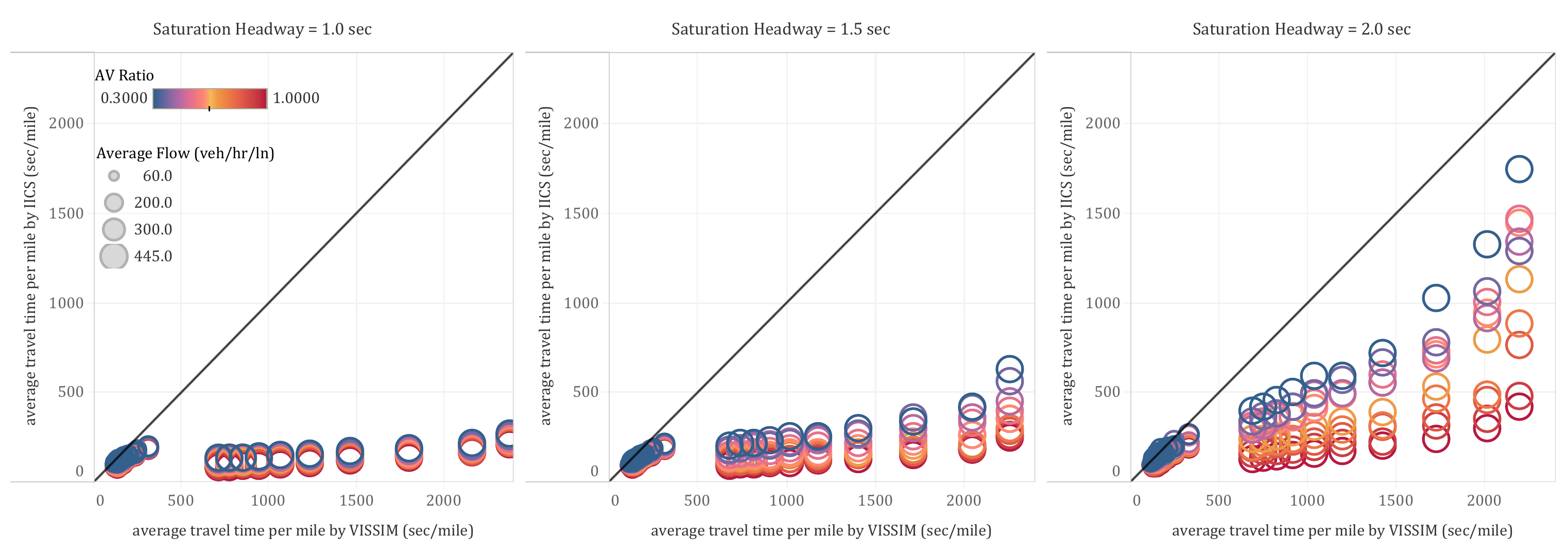}
\caption{Average travel time per mile compared between IICS and fully-actuated signal control for conventional vehicles in VISSIM.}\label{fig:MAT_VIS1}
\end{subfigure}
\centering
\begin{subfigure}{0.33\textwidth}
\centering
\includegraphics[width=1\linewidth]{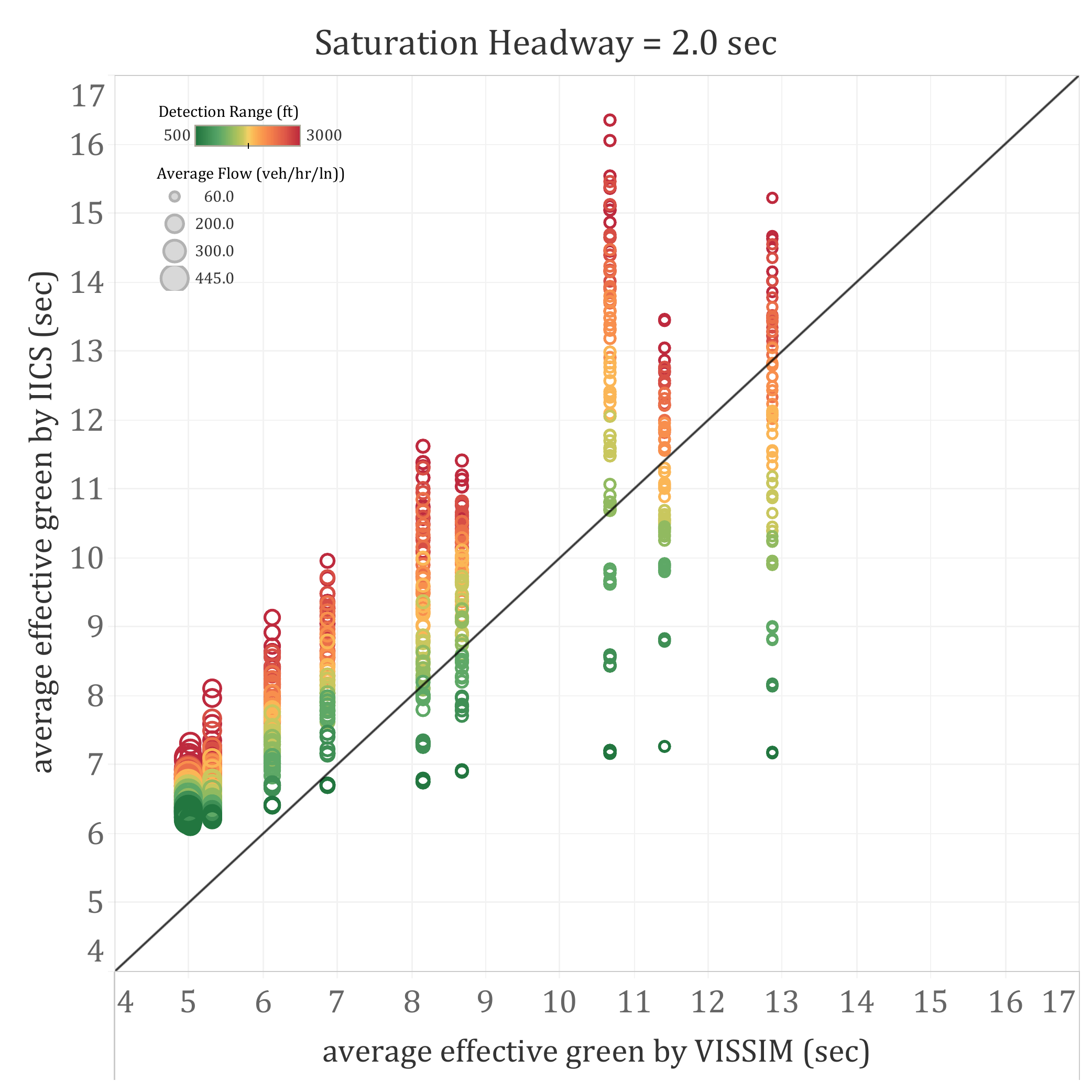}
\caption{Average effective green compared between IICS and fully-actuated signal control for conventional vehicles in VISSIM.}\label{fig:MAT_VIS2}
\end{subfigure} \caption{}\label{fig:VISMAT}
\end{figure}

\begin{figure}[hp]
\centering
\includegraphics[width=1\linewidth]{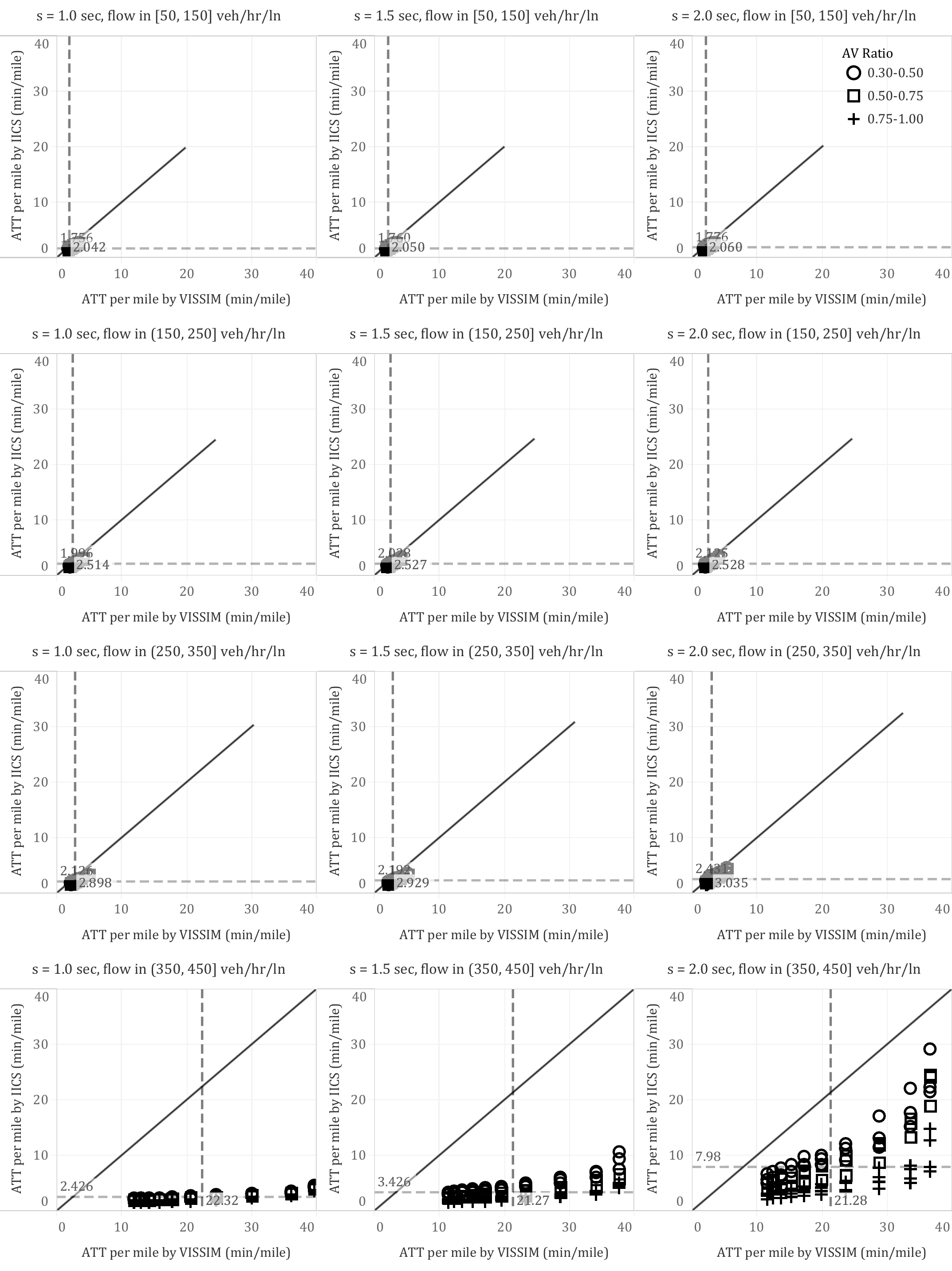}
\caption{Average Travel Time per mile compared between IICS and fully-actuated signal control for conventional vehicles in VISSIM (panels in same column associate with same saturation headway; Dashed lines show average values; Panels in same row associate with same flow spectrum; $s$ to be the saturation headway at the stop bar.)}\label{fig:VISSMAT3}
\end{figure}

Fig.~\ref{fig:MAT_VIS1} compares the average travel time per mile under IICS and fully actuated control strategy. The following can be observed: (1) The proposed IICS strategy leads to lower average travel times per mile compared to fully actuated control with all conventional vehicles (2) The rate of improvement increases as the saturation headway decreases, AV penetration rate increases, or average flow increases, Fig.~\ref{fig:VISSMAT3}. Our enhanced adaptive strategy made more frequent switches of right-of-way\textemdash using trajectories information\textemdash under higher flows, as shown in Fig.~\ref{fig:MAT_VIS2}.

\subsection{Practical Considerations for Implementation}
\label{S:4.3}
The term detection range refers to the farthest center-lane distance from each lane\textsc{\char13}s stop bar that roadside units sense a vehicle\textsc{\char13}s arrival. While present-day technology limits the maximum detection range, safety measures necessitate a minimum for the control algorithm to function in a timely manner. This section estimates the minimum detection range based on vehicle arrival information, deceleration capability, maximum crossing speed, and algorithm\textsc{\char13}s computation time.

\begin{figure}[h]
\begin{subfigure}{0.56\textwidth}
\centering
\includegraphics[width=1\linewidth]{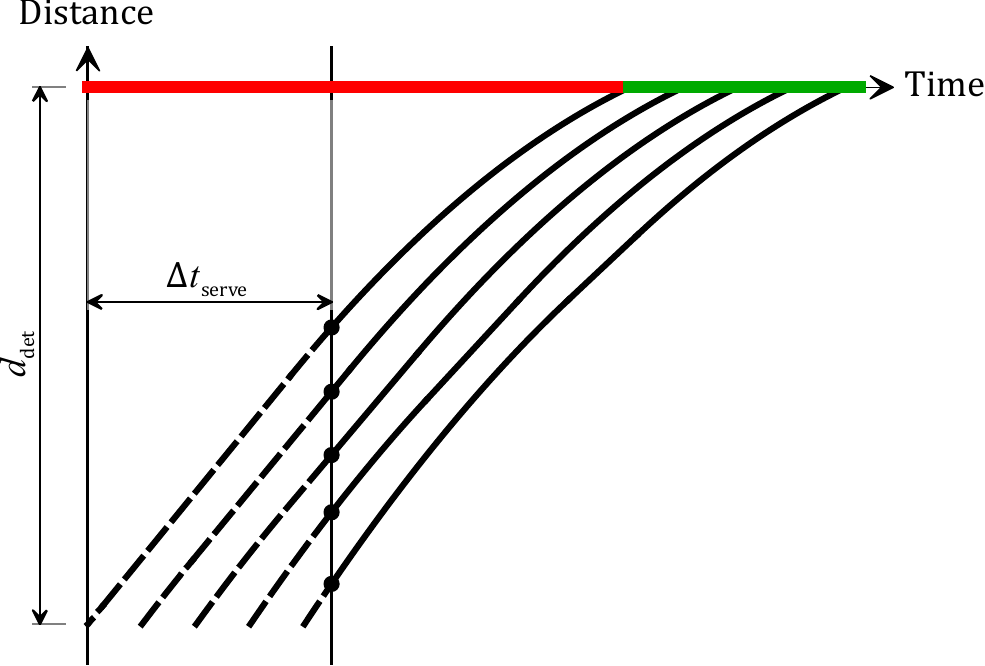}
\caption{The remaining distance to control the trajectory of an automated vehicle depends on initial speed, service time $\Delta t_{serve}$, detection distance $d_{det}$, deceleration rate, and maximum crossing speed at the stop bar.}\label{fig:late_traj_cost}
\end{subfigure}
\centering
\begin{subfigure}{0.43\textwidth}
\centering
\includegraphics[width=1\linewidth]{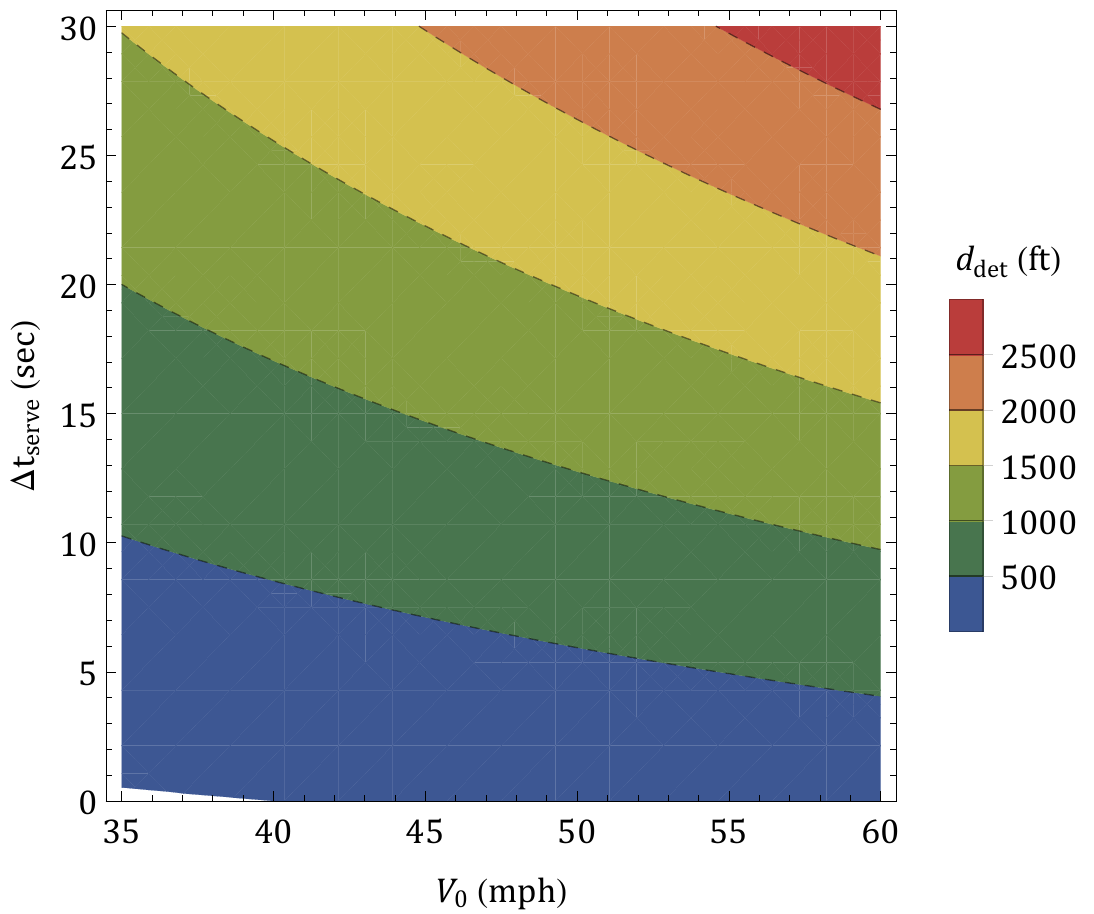}
\caption{Feasible detection range increases with service delay $\Delta t_{serve}$, and arrival speed $V_0$ (other parameters: deceleration rate is -15 ft/s$^2$, crossing speed is 40 mph).}\label{fig:Feas_det_range}
\end{subfigure} \caption{}\label{fig:communicationRestrict}
\end{figure}

We indicate serving time $\Delta t_{serve}$ as the time gap between detecting an automated vehicle and the time the vehicle is ready to follow a trajectory. Assuming the AV maintains constant speed and receives no trajectory, $\Delta t_{serve}$, it travels a portion of detection range based on the speed and serving time. After subtracting the uncontrolled traveled distance (Fig.~\ref{fig:late_traj_cost}), the remaining distance should be enough for the vehicle to safely decelerate. Eq.~\ref{eq:commRange} yields the minimum detection range to address the safe deceleration problem:

\begin{equation} \label{eq:commRange}
d_{det} \geq v_0 \Delta t_{serve}+\frac{v_c^2-v_0^2}{2 a^{max-}}
\end{equation}
Where:
\begin{enumerate}[leftmargin=1cm, labelsep=0.1cm, align=left, itemsep=-0.1cm, font=\small]
\item[$d_{det}$] the feasible detection range (in ft)
\item[$v_0$] the automated vehicle\textsc{\char13}s speed at the detection range (in ft/s)
\item[$v_c$] the maximum crossing speed at the stop bar (in ft/s)
\item[$a^{max-}$] the maximum deceleration rate (in ft/s$^2$)
\item[$\Delta t_{serve}$] the time delay to compute, and transmit AV\textsc{\char13}s trajectory.
\end{enumerate}

Eq.~\ref{eq:commRange} shows that the required detection range increases with initial speed and the time latency, see Fig.~\ref{fig:communicationRestrict}. For the current implementation of the proposed algorithm, the time latency is lower than a tenth of a second which makes it functional for very short detection ranges as low as 200 feet.

\section{Conclusions and Recommendations}
\label{S:5}
We developed a modeling framework (IICS) to integrate AVs with signalized intersection operations. The IICS model adjusts the signalization and trajectory of automated vehicles on a real-time basis to minimize total travel time delay. The method incorporates the following factors into the decision-making process:

\begin{itemize}[leftmargin=1cm, labelsep=0.1cm, align=left, itemsep=-0.1cm, font=\small]
\item Vehicle arrival information
	\begin{itemize}[leftmargin=1cm, labelsep=0.1cm, align=left, itemsep=-0.1cm, font=\small]
	\item Traffic stream composition (e.g. automated versus conventional ratio)
	\item Individual vehicle
    	\begin{itemize}
    	\item Spatial information (e.g. speed and location of an entering vehicle)
        \item Attributes (e.g. acceleration/deceleration capabilities)
    	\end{itemize}
        \item Flow fluctuation (e.g. randomness of incoming vehicles in each lane)
	\end{itemize}
    \item Phase plan (e.g. Serving maximum movements without merge or crossing conflicts)
    \item Speed limits (e.g. the maximum allowable speed near an intersection, and the maximum crossing speed at the stop bars)
\end{itemize}

The proposed algorithm prevented queue formation for undersaturated conditions, jointly optimized signalization and trajectories of automated vehicles, and utilized both connectivity and programmability of AVs. The implementation also made decisions in a fraction of a second suitable for real-time application.

Several assumptions limit the scope of this study and left unanswered questions for future research. The process turns unstable for oversaturated conditions as queue formation becomes inevitable. The situation requires a more rigorous signal control logic that considers residual vehicles at the end of red intervals.

\section*{Acknowledgment}
This material is based upon work supported by the National Science Foundation (under Grant No. 1446813, titled: Traffic Signal Control with Connected and Autonomous Vehicles in the Traffic Stream).


\section*{References}








\bibliographystyle{model5-names}\biboptions{authoryear}



\bibliography{sample}







\end{document}